\documentclass[a4paper,11pt]{article}
\pdfoutput=1 

\usepackage{jinstpub} 

\usepackage{lineno}

\title{\boldmath  LAPPD operation using ToFPETv2 PETSYS ASIC}

\author[a,1]{A. Seljak\note{Corresponding author.}}
\author[c,a]{M. Bra\v{c}ko}
\author[b,a]{R. Dolenec}
\author[b,a]{P. Kri\v{z}an}
\author[a]{A. Lozar}
\author[a]{R. Pestotnik}
\author [c,a]{and S. Korpar}

\affiliation[a]{Experimental particle physics department - F9, Jozef Stefan Institute,\\ Ljubljana, Slovenia}
\affiliation[b]{University of Ljubljana, Faculty of Mathematics and Physics,\\Ljubljana, Slovenia}
\affiliation[c]{University of Maribor, Faculty of Chemistry and Chemical Engineering\\Maribor, Slovenia}

\emailAdd{Andrej.Seljak@ijs.si}

\abstract {Single photon sensitive detectors used in high energy physics are, in some applications, required to cover areas the size of several m$^2$, and more specifically in very strong demand with an ever finer imaging and timing capability for Cherenkov Ring Imaging Detector (RICH) configurations. We are evaluating the Large Area Picosecond Photo-detector (LAPPD) produced by INCOM company, as a possible candidate for future RICH detector upgrades. In this work we perform tests on the second generation device, which is capacitively coupled to a custom designed anode back plane, consisting of various pixels and strips varying in size, that allows for connecting various readout systems such as standard laboratory equipment, as well as the TOFPET2 ASIC from PETsys company. Our aim is to  evaluate  what can be achieved by merging currently available technology, in order to find directions for future developments adapted for specific uses.}

\keywords{Photon detectors for UV, visible and IR photons (vacuum) (photomultipliers, HPDs, others), Data acquisition concepts, Data acquisition circuits }

\proceeding{2022 Topical Workshop on Electronics for Particle Physics\\
	19$^{\text{th}}$ till 23$^{\text{th}}$ September\\
	Bergen, Norway}

\begin{document}
\maketitle
\flushbottom

\section{Introduction and motivation}
\label{sec:intro}

\begin{figure}[htbp]
	\centering 

	\includegraphics[width=.97\textwidth]{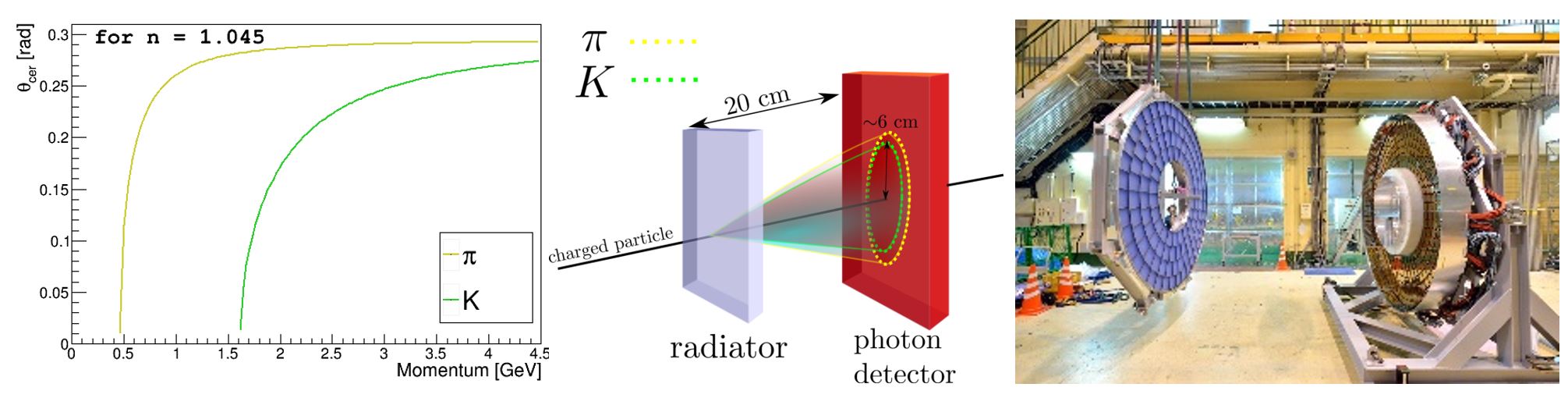}
	\caption{\label{fig:one} Kaon and pion Cherenkov angle emission as a function of particle momentum. RICH detector configuration using Aerogel as radiator, and image of the Belle II ARICH detector installed at Belle II spectrometer.}
\end{figure}

Ring Imaging Cherenkov detectors represent an indispensable tool for particle identification at particle colliding accelerators. A small number of photons being emitted by charged particle tracks in  radiators, such as aerogels or gas mixtures, need to be detected on a detector plane of usually very large size, and require sub cm pixel pitch resolution for imaging Cherenkov rings. To give a sense of the covered area, the Belle II ARICH photon detector~\cite{ARICH} (Figure~\ref{fig:one}) has an area of around 4.4 m$^2$, while the RICH1 and RICH2 photon area detectors in LHCb is around 1.2 m$^2$ each \cite{TDR}.  Considering the target luminosity of 2×10$^{33}$cm$^{-2}$s$^{-1}$  ~\cite{floris} for the High Lumi LHC upgrade, an ever finer pixel granularity of the  detector is needed to keep its occupancy within an expected few percent level.  This creates a considerable challenge in the amount of channels to be read at  bunch crossing frequencies, which in turn creates a real problem in  data transmission rates. Considering a binary readout of a detector the size of a m$^2$, the pixel pitch of a mm, at nominal 40 MHz bunch crossing rate, yields a raw data rate of 40 Tbits/s.  Nowadays the time of flight (TOF) information from the vertex point to the detector plane is also under consideration. The TOF information below 10~ps would in general improve charged particles identification, so it is strongly considered for future upgrades ~\cite{floris}. In such case, the estimates for data rates increment further to encapsulate fine timing information.

The development of recent MCP based single photon sensitive imaging detectors has found solutions for covering large areas using several mm sized pixels \cite{drs}, and in an opposite extreme of having pixelated chips with pixel pitch in the hundreds micrometer range \cite{timepix}, as well as trade off solutions using crossed strip pattern readout schemes \cite{SSL}.  All of these approaches are scalable, although the fine pitch pixelated electronics have a large power consumption,  are unpractical for very large surfaces, and the provided granularity overreaches RICH applications. The crossed strip solution has position readout ambiguity issues, and the pixelated variant requires a lot of readout channels. 

The mm pixel pitch seems to fall in a technological gap where pixelated chip based implementations are impractical and too fine pitched, binary readout pixels of several mm resolution is too coarse, and timing dependent pixel position estimation schemes are inadequate due to count rate inability. Hence some 3 dimensional stacking of a pixelated anode pads, and a wire bonded or ball bonded  ASIC dies on an interpose board represents a viable solution. Although at this point in time it isn't yet clear which kind of photo detector will be favored for a particular future RICH system,  in this work we analyze a present state of the art detector and readout electronics, which were not specifically designed to work in conjunction, but represent a doable starting point for future developments.

\section{ Large Area Picosecond Photo-Detector }
\label{sec:LAPPD}

\begin{figure}[htbp!] 
	\centering 
	
	\includegraphics[width=.4\textwidth]{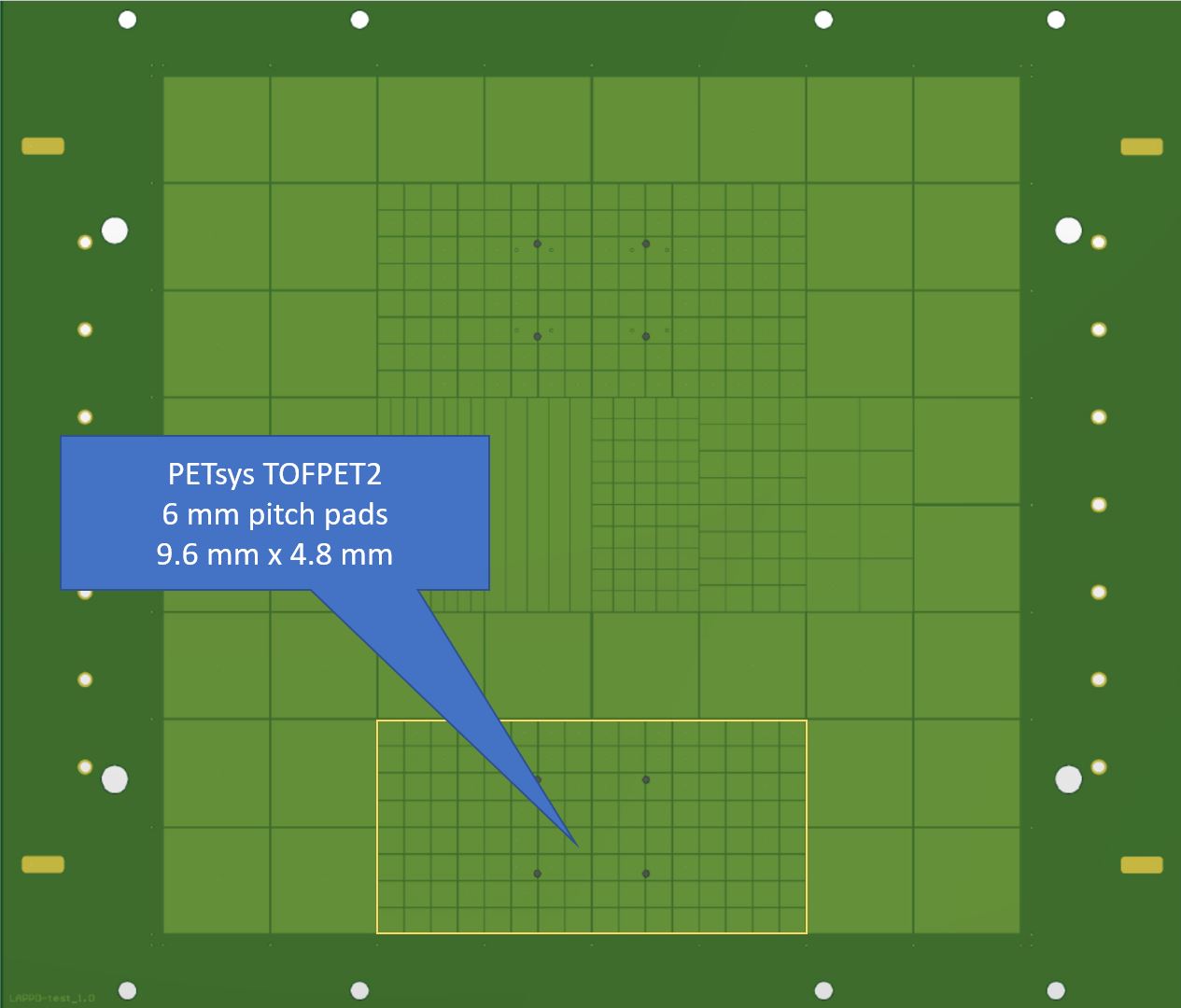}
	\includegraphics[width=.52\textwidth]{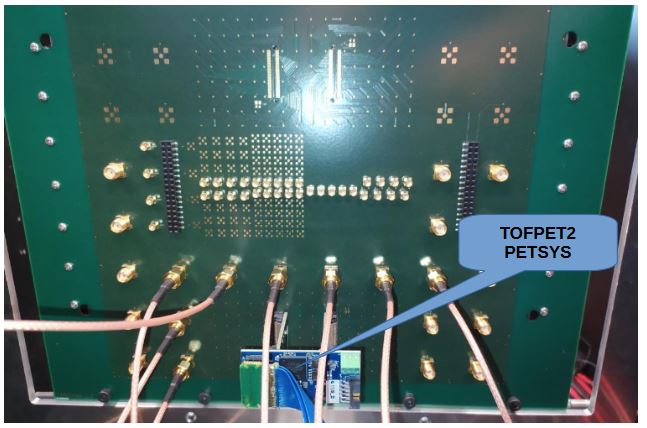}

	\caption{\label{fig:two} Screenshot of the PCB drawing showing various pad sizes and their locations, image of the PCB board attached to the LAPPD detector and the location of the 128 channel PETsys readout.}
\end{figure}

The LAPPD Generation II  ~\cite{lappd}  is the size of 8x8 inch, which uses a photo-cathode on the entrance window for quantum conversion, and a dual chevron stacked 20 $\mu$m Micro Channel Plate (MCP) for signal multiplication. The gain at nominal power supply voltages is around 10$^7$, while the dark count self-contribution is ~150~counts/s/cm$^2$.  These vacuum sealed devices enable to capacitively couple an anode plane through a 4 mm thick glass on the back side, thus enabling the user to design a customized anode plane,  like we designed ours for various tests  (Figure \ref{fig:two}),  and using the readout system of one’s preference.  Some commercially designed readout systems using waveform sampling technology to sense the strip line patterns on the anode backplane, extrapolating the X and Y coordinate from the time of arrival at each end in one dimension, and charge sharing for the other were already proposed  ~\cite{drs}. We explore the possibility to physically pixelate the device in order to increment the count rate ability, as well to improve on the spatial resolution compared to the strip readout (0.7~mm~x~6.1~mm~\cite{drs}). The custom designed PCB based  back-plane   (Figure \ref{fig:two}) is segmented with various pixel sizes ranging from inch by inch, half inch pitch, to mm sized strips, as well as 6 mm pitch pads.

\begin{figure}[htbp!]
	\centering 
	
	\includegraphics[width=.44\textwidth]{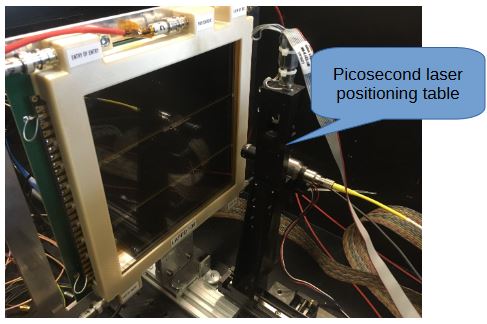}
	\includegraphics[width=.31\textwidth]{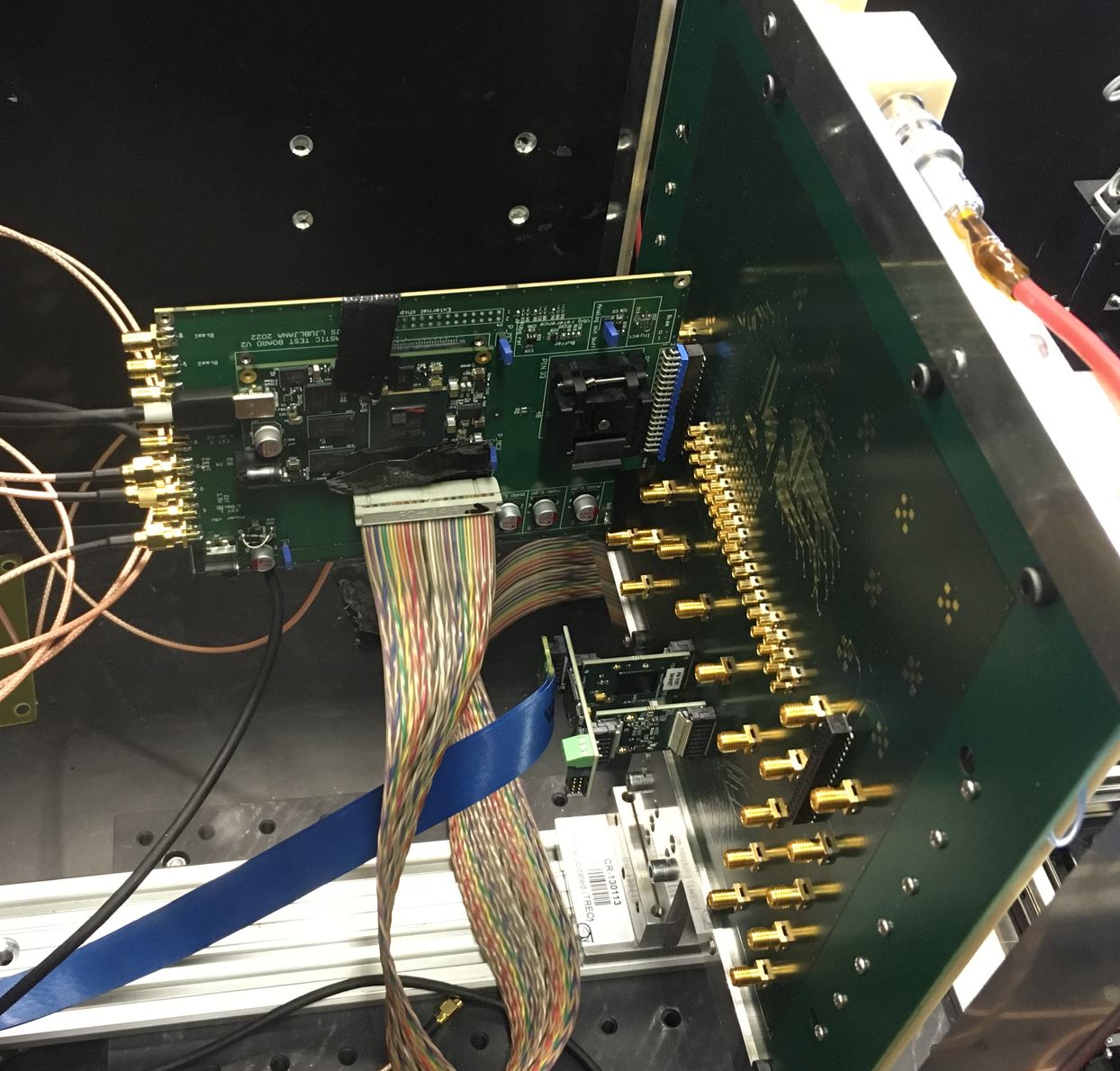}
	
	\caption{\label{fig:three} Image of the LAPPD detector in the setup, image of the anode plane with FASTic and PETsys attached electronics.}
\end{figure}

Our setup is composed of the LAPPD photo detector, the ALPHALAS 20 ps (FWHM) laser ~\cite{alphalas} attenuated to mostly detect single photon electron, having a focusing head mounted on translation stages in order to position the 100 	$\mu$m laser spot, and an oscilloscope for monitoring signals (Figure~\ref{fig:three}). We observed that, due to the thickness of the glass  backplane, the induced charge spreads considerably. If one would look into making a binary readout, the minimal pixel pitch for this particular device is of 6 mm to collect some 30 percent of charge per event. Using 6 mm pads, one PETsys readout module (128 channels) covers 46 cm$^2$, as shown in figure~\ref{fig:two}~on~the~left. With this system we measure the jitter between firing the laser and the photon time of arrival (TOA),  as well as an approximation of charge for every pixel.  

\section{ PETsys TOFPET2 ASIC and readout} 
\label{sec:PETSYS}

\begin{figure}[htbp]
	\centering 
	
	\includegraphics[width=.30\textwidth]{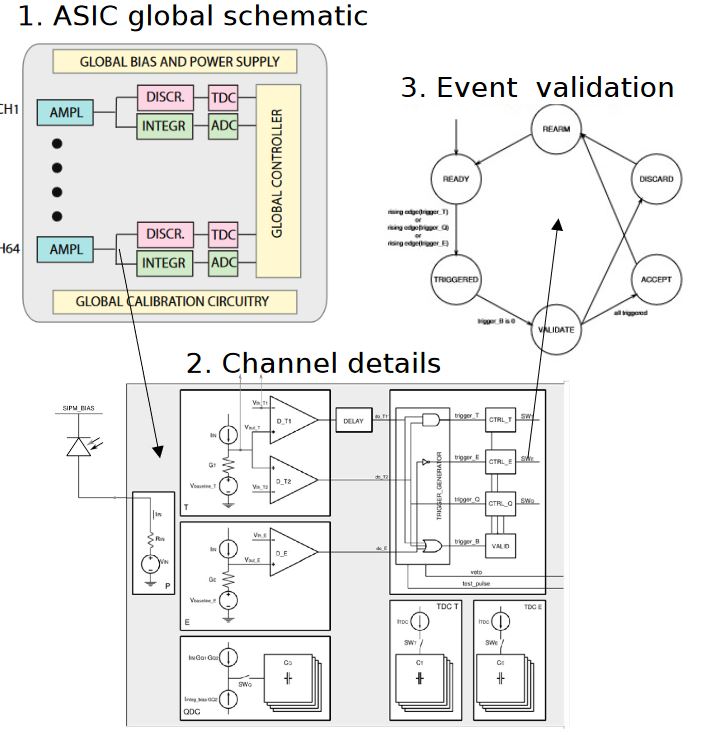}
	\includegraphics[width=.64\textwidth]{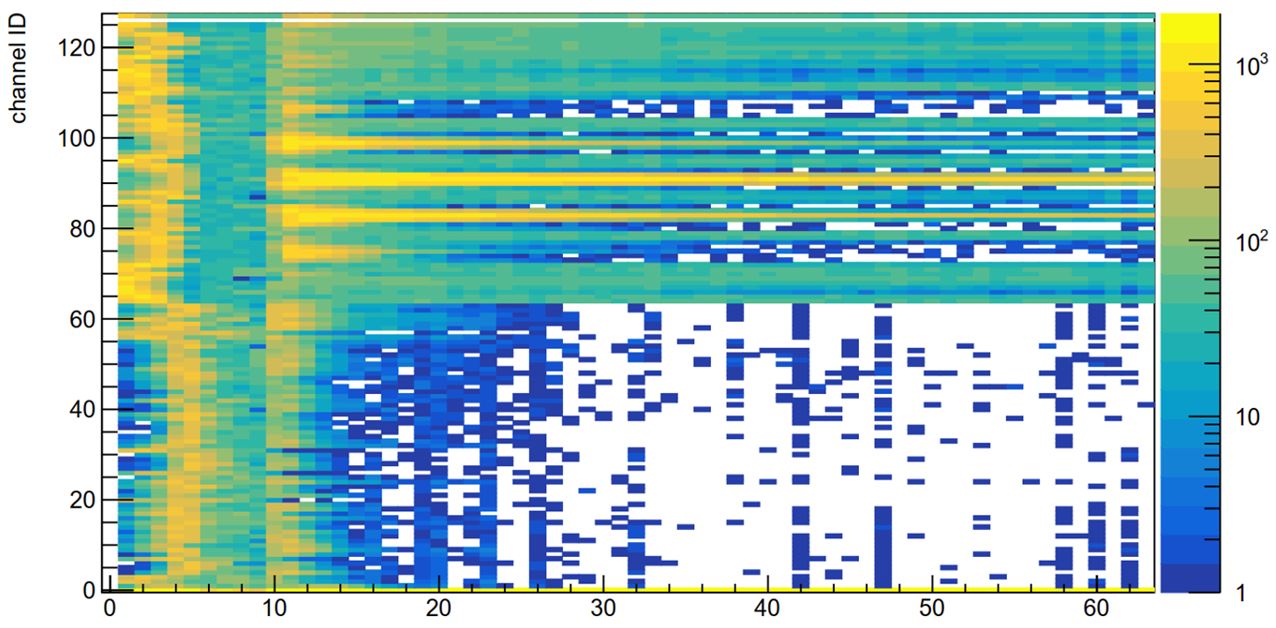}

	\caption{\label{fig:four}  Block schematic of the ASIC - left, calibrated threshold scan with channel 91 illuminated (Y-channel, X-Threshold, Z-Events) - right}
\end{figure}

The PETsys TOFPET2 ASIC ~\cite{petsys} is provided in a contained module having two chips, 64 channels each, as depicted on figure~\ref{fig:two}~on~the~right. Such modules  aggregate on an FPGA based board  providing up to 1024 readout channels, being read from a PC via an ethernet cable.  The chip's channel is composed of an input signal front end, a baseline and threshold adjustable level toward a comparator, firing internal TDCs  providing the TOA and time over threshold (TOT) information. Since the chip was conceived for SiPM/crystal PET, its energy output information is not sensitive enough to measure few hundred~$f$C MCP signals, however those can be derived from the TOT information with limited accuracy. The  TOT information can be obtained instead of the energy measurement via the chips settings. The internal TDC bin is  30 ps while the dynamic range for measuring charge is up to 1500 pC. Our module version allows for input pulse polarity reversal, and the maximum hit rate per channel is 600 kHz.  Figure \ref{fig:four}  left shows a schematic view  of the ASIC's channel construction ~\cite{petsys}.  We first calibrated the system by performing a threshold scan and adjusted the channel baselines.  For the threshold scan shown in figure \ref{fig:four} we illuminated multiple times near the center of the pad connected to the arbitrary chosen channel (91). The intensity of the light pulses was adjusted to detect mostly single photo electron signals. This picture shows that adjacent pixels sense the charge spread as well. Using the information provided by the chip's TDC, the TOA jitter was found to be around 200 ps sigma.

\begin{figure}[htbp]
	\centering 
	
	\includegraphics[width=.9\textwidth]{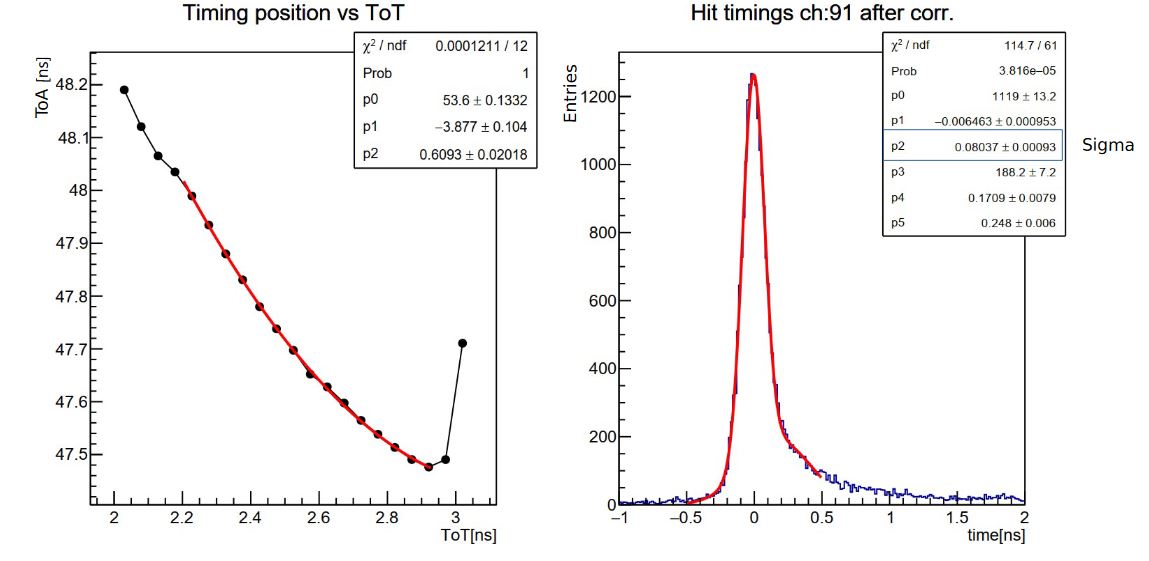}

	\caption{\label{fig:five}  TOA as a function of TOT, and time walk corrected TOA for channel 91 with the result of 80 ps sigma from a double Gaussian fit.}
\end{figure}

Analyzing the same set of data, we explore the relation between the TOA and TOT in search for a possible adjustment of the time walk (Figure~\ref{fig:five}), due to the fluctuation in the amount of charge for a single photon coming from the MCP. It turns out the chip's sensitivity level  in TOT mode is capable to provide enough information to correct the TOA, and we managed to obtain the timing resolution of the system down to 80 ps sigma, using the information from the illuminated channel. Such figure could probably be improved considering the information from the neighboring channels.  

\begin{figure}[htbp]
	\centering 
	
	\includegraphics[width=.7\textwidth]{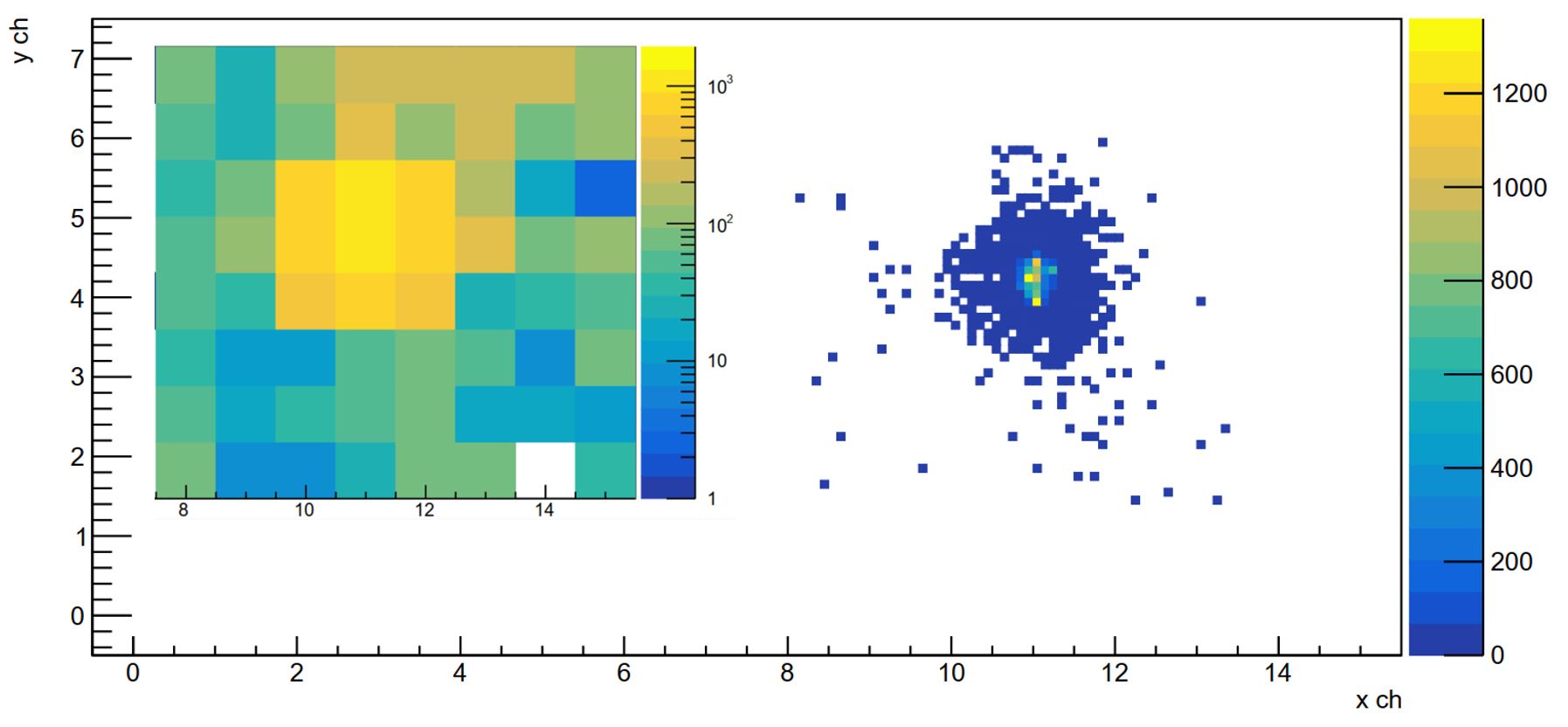}

	\caption{\label{fig:six}  Left top is the integrated response by illuminating channel 91. One unit on X and Y scale is equivalent to 6 mm.  The recalculated entrance position using the center of gravity estimation for every event improves the spatial resolution to around millimeter FWHM, as presented on the right side of the plot for the same data.   }
\end{figure}

We analyze the possibility to  improve the photon entrance position information below the pad physical (6 mm)  size.  The hit count map was produced by histograming events (Figure~\ref{fig:six}), while illuminating the center of a single pixel, and with a fixed threshold. We used the TOT information of the ASIC to estimate the
charge amount and calculate the photon entrance position from the charge center of gravity for every event. The approach favorably improved the spatial resolution, which we found to be in mm range Full Width Half Maximum (FWHM).

\section{Conclusion}
\label{sec:disc}

The LAPPD is a very large device with the flexibility of making custom pads to attach the readout system. During tests we noticed that due to charge spread in the enclosure back-plane, the minimal anode pad pitch is around 6~mm to perform a binary readout.   Below this size one can employ charge center of gravity reconstruction for a given event, in order to achieve a better spatial resolution. The Petsys - ToFPET2  is a commercially available ASIC, available along with a  complete readout system, which seems to work quite well with the LAPPD. The ASICs front end seems to be capable of resolving fC charges with sub 100 ps timing resolution, and enough granularity to provide information in order to enhance the spatial resolution. Overall it is possible to achieve a mm size FWHM  pixels, with a timing blow 100~ps. Furthermore the system has 1024 channels which, using a pad size equal or above 0.9 cm, can handle two LAPPD devices.  Perhaps in some future chip redesign  the levels and functionality will be addressed to increment the count rate as well. The evaluated combination naturally isn’t yet capable to handle event rates of 40 MHz, however it enabled a demonstration of excellent resolution with fine timing abilities, where large surfaces need to be covered and the expected single photon occurrences are very low.

All authors declare that they have no known conflicts of interest in terms of competing financial interests or personal relationships that could have an influence or are relevant to the work reported in this paper.  The authors acknowledge the financial support from the Slovenian Research Agency (research core fundings No. P1-0389 and P1-0135, postgraduate research funding and project No. J2-1735).

\end{document}